\documentclass[aps,twocolumn,epsfig,prl,showpacs]{revtex4}

\usepackage{amsmath}
\usepackage{graphicx}

\def\nn{\nonumber}

\begin{document}

\title{Theory of superconductivity of carbon nanotubes and graphene}

\author{K. Sasaki}
\email[Email address: ]{sasaken@flex.phys.tohoku.ac.jp}
\affiliation{Department of Physics, Tohoku University and CREST, JST,
Sendai 980-8578, Japan}

\author{J. Jiang}
\altaffiliation[Present address: ]{Department of Physics, North Carolina
State University, Raleigh, North Carolina 27695, USA}
\affiliation{Department of Physics, Tohoku University and CREST, JST,
Sendai 980-8578, Japan}

\author{R. Saito}
\affiliation{Department of Physics, Tohoku University and CREST, JST,
Sendai 980-8578, Japan}

\author{S. Onari}
\affiliation{Department of Applied Physics, Nagoya University,
Nagoya 464-8603, Japan}

\author{Y. Tanaka}
\affiliation{Department of Applied Physics, Nagoya University,
Nagoya 464-8603, Japan}

\date{\today}
 
\begin{abstract}
 We present a new mechanism of carbon nanotube superconductivity 
 that originates from edge states which are specific to graphene.
 Using on-site and boundary deformation potentials
 which do not cause bulk superconductivity, 
 we obtain an appreciable transition temperature for the edge state.
 As a consequence,
 a metallic zigzag carbon nanotube having open boundaries
 can be regarded as a natural superconductor/normal metal/superconductor
 junction system, 
 in which superconducting states are developed locally at both ends of
 the nanotube and 
 a normal metal exists in the middle.
 In this case, a signal of the edge state superconductivity appears as
 the Josephson current which is sensitive to the length of a
 nanotube and the position of the Fermi energy.
 Such a dependence distinguishs edge state superconductivity 
 from bulk superconductivity.
\end{abstract}


\maketitle

Superconductivity in carbon nanotubes (NTs) 
has been attracting much attention
due to its high superconducting transition temperature, 
$T_{\rm c} \agt 10$ K.~\cite{tang01,takesue06}
However, it is well-known that 
superconductivity in low dimensional (quasi-1D) systems
is difficult to produce due to 
low density of states (DOS)~\cite{saito98book},
strong quantum fluctuations and other phenomena in such systems.
Moreover, metallic NTs exhibit
ballistic transport properties at low
temperatures,~\cite{bachtold00}
which suggests a weak electron-phonon (el-ph) interaction 
for the conducting electrons.
It is surprising that superconductivity is realized in NTs 
at such high values of $T_{\rm c}$.
The mechanism of NT superconductivity
is a critical issue and determining it will be a 
valuable contribution not only to NT science 
but also to nanotechnology.

Superconductivity has been observed in different types of NTs.
Tang {\it et al}. reported $T_{\rm c} \sim 15$ K 
for single-wall NTs (SWNTs) having a diameter 
of 0.4 nm.~\cite{tang01} 
Takesue {\it et al.}  
found an abrupt drop in the zero-bias resistance
at 12 K for multi-wall NTs (MWNTs) having an outer diameter
of $\sim$ 10 nm.~\cite{takesue06}
It is not straightforward
to explain the results obtained in experiments.
For instance,
the DOS at the Fermi energy of 
a SWNT appears to be too small to give rise to such high $T_{\rm c}$.
Kamide {\it et al}.~\cite{kamide03} and
Barnett {\it et al}.~\cite{barnett05} 
considered that curvature of $(5,0)$ SWNTs
may increase the DOS.
However, a large DOS may induce charge density wave (CDW) 
before superconductivity occurs. 
Conn\'etable {\it et al}. showed that 
$(5,0)$ and $(3,3)$ SWNTs undergo a CDW transition 
at temperatures above room temperature.~\cite{connetable05} 
Thus, small diameter SWNTs may be insulators.
Moreover the curvature effect is negligible for MWNTs.
The origin of NTs superconductivity
can not be explained by curvature-induced DOS and 
a new explanation is needed.

Here, we focus our attention 
on the large local DOS (LDOS) given by edge states
which are intrinsic to graphene.
The edge states are electronic localized states
that exist around the zigzag edge of graphene and a
SWNT.~\cite{fujita96,nakada96}
The energy dispersion of edge states 
is located near the Fermi energy ($E_{\rm F}=0$).
The value of LDOS depends on 
the energy bandwidth ($W$) of the edge states. 
Recent experiments involving 
scanning tunneling microscopy/spectroscopy
(STM/STS) at the zigzag edge of
graphene,~\cite{niimi05,kobayashi05} and 
angle-resolved photo-emission spectroscopy (ARPES) of Kish
graphite,~\cite{sugawara06} 
showed that the edge states are
located below the Fermi energy and have a finite $W$.
Although edge states of NTs 
have not been observed so far, 
it is possible to consider the edge states of a SWNT 
with a zigzag edge
as well as those of a graphene sheet.

In this letter, 
we calculate $T_{\rm c}$ as a function of $W$ and $E_{\rm F}$,
and obtain an appreciable values for $T_{\rm c}$
of the edge states of zigzag SWNTs and graphene.
As a result, we predict that the superconductivity of a SWNT 
is given by a natural superconductor/normal metal/superconductor
junction (SNS) system, 
in which superconducting states develop locally at both ends of
the SWNT and a normal, ballistic state exists 
in the middle of the SWNT. 
Remarkably, the bulk part of a SWNT need not be superconducting
since Josephson supercurrent flows in the middle 
as a result of the proximity effect
when the superconducting edge states have different phases at both
ends.
We note that proximity-induced supercurrents have been observed in 
Ta/SWNTs/Au~\cite{kasumov99}
and Nb/MWNTs/Al systems.~\cite{haruyama04}
The Josephson current of a metallic zigzag SWNT 
depends on the length ($L$) and temperature ($T$).
The amplitude of the current is proportional to 
$\exp(-L/\xi_N)$
when $T < T_0$
where $T_0 \sim 20 \mu {\rm m}/L$ K and
$\xi_N \sim 10^3 {\rm K}/T $ nm is
the coherence length,~\cite{wakabayashi02}
which is a characteristic feature of conventional SNS transport
theory for the clean limit.~\cite{kulik70}
A length dependence of the current
distinguishes edge-state superconductivity 
from bulk superconductivity.

The edge-state superconductivity 
has the following advantages in explaining the
experiment performed by Takesue {\it et al.},~\cite{takesue06}
(1) the edge states are robust against static surface deformation
which is relevant for CDW instability,~\cite{fujita97,sasaki06jpsj}
(2) the el-ph interaction for the edge states 
is strong compared with that for delocalized states, and
(3) $T_{\rm c}$ is sensitive to $W$
and the energy position of $E_{\rm F}$,
which are all consistent with the fact
that the superconductivity is sensitive to the 
junction structures of the Au electrode/MWNTs.~\cite{takesue06}
Enhancement of $T_{\rm c}$ at the edge is important for
understanding superconductivity of a general surface state,
not only for graphite materials but also for 
noble-metals such as gold.~\cite{kevan87}

\begin{figure*}[htbp]
 \begin{center}
  \includegraphics[scale=0.85]{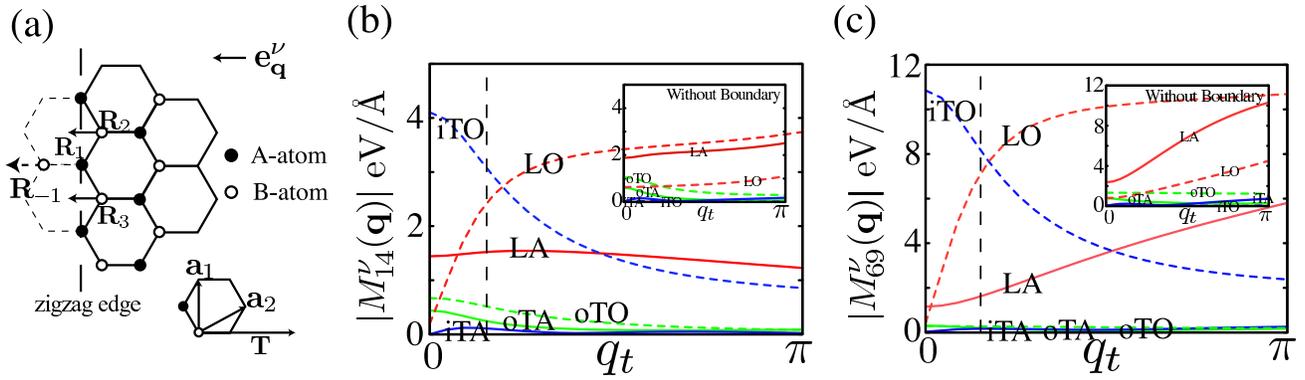}
 \end{center}
 \caption{(Color online)
 (a)
 Carbon atoms at ${\bf R}_{p>0}$ near the zigzag edge,
 and a fictitious carbon atom at ${\bf R}_{-1}$
 are illustrated to show an enhancement of the 
 on-site deformation potential at the boundary.
 The boundary deformation potential
 is large for optical modes whose ${\bf e}_{\bf q}^\nu$ is 
 parallel to ${\bf T}$.
 (b,c) $|M_{ij}^\nu({\bf q})|$
 of the $(60,0)$ SWNT with
 (b) $k(1)(=7\pi/10) \to k(4)(=4\pi/5)$ and 
 (c) $k(6)(=13\pi/15) \to k(9)(=29\pi/30)$
 are plotted as a function of $q_t$ where $q=\pi/10$.
 The inset is the matrix element including (fictitious) 
 carbon atoms at ${\bf R}_{p<0}$. 
 Three solid/dashed curves represent acoustic/optical phonon modes:
 oTA/oTO(green), iTA/iTO(blue) and LA/LO (red).
 The vertical dashed lines represent $q_t=\sqrt{3}q$
 ($\sqrt{3} = |{\bf T}|/|{\bf a}_1|$).
 } 
 \label{fig:me}
\end{figure*}

The edge states are zero-energy ($E(k)=0$) eigenstates of
the nearest-neighbor (nn) tight-binding Hamiltonian,
${\cal H}_{\rm nn} | \Psi(k) \rangle = E(k) | \Psi(k)
\rangle$.~\cite{fujita96,sasaki05prb,sasaki06apl}
$k \equiv {\bf k} \cdot {\bf a}_1$
is the wavevector around the tube axis
where ${\bf a}_1$ is the unit vector along the edge
(see Fig.~\ref{fig:me}(a)).
The edge states exist for $2\pi/3 < k < 4\pi/3$.
The wavefunction is written as
\begin{align}
 | \Psi(k) \rangle
 = \sum_{i \in {\rm A}} C_i(k)|\phi({\bf R}_i) \rangle,
 \label{eq:wf_edge}
\end{align}
where $|\phi({\bf R}_i) \rangle$ is the 2${\rm p}_z$ orbital,
and $C_i(k)$ is the amplitude at ${\bf R}_i$
and has a value on one of the two sublattices (A and B)
of graphite.~\cite{fujita96}
In the direction of the SWNT axis,
the magnitude of $C_i(k)$ quickly decays from the edge
to the interior region.
The localization length is given by
$\xi(k) = -|{\bf T}|/2\ln |2 \cos(k/2)|$ ($2\pi/3<k<4\pi/3$) 
where
${\bf T} = 2{\bf a}_2 - {\bf a}_1$ is the translation
vector.~\cite{sasaki06apl,sasaki05prb}
When we incorporate the next nearest-neighbor (nnn) 
transfer integral $\gamma_n$ into the Hamiltonian, 
the energy dispersion of the edge states becomes
$E(k) = \gamma_n (2\cos k +1)$ ($2\pi/3<k<4\pi/3$)
where the value of $\gamma_n=0.3$ eV is adopted.~\cite{sasaki06apl,porezag95}
The calculated results explain the
STS~\cite{niimi05,kobayashi05} and
ARPES~\cite{sugawara06} experiments.
Hereafter we treat $W$ ($=\gamma_n$) and the position of the Fermi
energy as independent parameters.

The el-ph interaction for the edge states 
shows a different behavior from that for delocalized states. 
The el-ph interaction consists of on-site
and off-site deformation potentials.~\cite{jiang05}
It is pointed out that, 
for a backward scattering of delocalized states,
the on-site deformation potentials on two sublattices
cancel with each other due to a phase difference 
of the wavefunction at the two sublattices.~\cite{suzuura02}
This is a reason 
why metallic NTs show a ballistic transport property.
However, 
the cancellation of the on-site deformation potential 
does not work for the edge states
since the wavefunction of the edge state has
an amplitude only on one of the two sublattices.
Furthermore, because of a lack of translational symmetry at the edge,
a strong el-ph interaction for optical phonon modes
is expected for the edge.
Thus the understanding of the el-ph interaction for the edge states 
is essential for the present problem.

The el-ph Hamiltonian is defined by
${\cal H}={\cal H}_0 + {\cal H}_{\rm int}$,
where ${\cal H}_0$ represents the edge states and phonon,
and ${\cal H}_{\rm int}$ is the el-ph interaction.
${\cal H}_0$ is given by
$\sum_{k} E(k) c_k^\dagger c_k
+ \sum_{{\bf q},\nu} 
\omega_\nu({\bf q}) b_{{\bf q},\nu}^\dagger b_{{\bf q},\nu}$,
where $c_k$ is the annihilation operator of edge state and
$b_{{\bf q},\nu}$ is the annihilation operator
of $\nu$-th phonon mode with momentum ${\bf q}$ and energy 
$\omega_\nu({\bf q})$.
For a graphite unit cell,
there are six phonon eigen-modes;
out-of-plane tangential acoustic/optical mode (oTA/oTO),
in-plane tangential acoustic/optical mode (iTA/iTO), and
longitudinal acoustic/optical mode (LA/LO).
$\omega_\nu({\bf q})$ and phonon eigenvector
are obtained by solving 
a $6\times 6$ dynamical matrix.~\cite{jiang05}
The el-ph interaction is given by
\begin{align}
 {\cal H}_{\rm int} =
 \frac{1}{\sqrt{N_{\rm u}}} \sum_{k,k'}
 \sum_{q_t,\nu}
 \alpha^\nu_{k k'}({\bf q}) 
 (b_{{\bf q},\nu}+b_{{\bf -q},\nu}^\dagger)
 c_{k'}^\dagger c_k,
 \label{eq:Hint}
\end{align}
where $N_{\rm u}$ is the number of graphite unit cells in a SWNT, and
$\alpha^\nu_{kk'}({\bf q})$ is the el-ph coupling connecting two edge
states $k$ and $k'$ by $\nu$-th phonon mode with momentum ${\bf q}$.
Due to the momentum conservation along the edge,
$k'=k + q$ ($q\equiv {\bf q}\cdot {\bf a}_1$),
while the wavevector perpendicular to the edge
$q_t$ ($\equiv {\bf q}\cdot {\bf T}$) is needed to sum over the
Brillouin zone.

We calculate $\alpha^\nu_{kk'}({\bf q})$ using the deformation
potential,
$\delta V = -\sum_p \nabla v({\bf R}_p) \cdot {\bf u}({\bf R}_p)$,
where ${\bf u}({\bf R}_p)$ is the displacement vector and 
$v({\bf R}_p)$ is the pseudo-potential 
of a carbon atom at ${\bf R}_p$.
The present pseudo-potential is used for
calculating resonance Raman intensity in which the calculated results
explain chirality and diameter dependence of Raman
intensity quantitatively.~\cite{jiang05} 
${\bf u}({\bf R}_p)$ can be expanded 
by phonon normal modes as
${\bf u}({\bf R}_p) =
\sum_{{\bf q},\nu} (A^\nu({\bf q})/\sqrt{2N_{\rm u}})
( b_{{\bf q},\nu}+b_{{\bf -q},\nu}^\dagger )
{\bf e}^\nu_{\bf q}({\bf R}_p)
e^{i{\bf q}\cdot {\bf R}_p}$,
where ${\bf e}^\nu_{\bf q}({\bf R}_p)$ is the normalized 
eigenvector at ${\bf R}_p$ and 
$A^\nu({\bf q})=\hbar/\sqrt{m_c \omega_\nu({\bf q})}$ is 
the phonon amplitude.
From $\langle \Psi(k') | \delta V | \Psi(k) \rangle$,
we obtain
$\alpha^\nu_{kk'}({\bf q})
\equiv A^\nu({\bf q}) M^\nu_{kk'}({\bf q})/\sqrt{2}$,
where $M^\nu_{kk'}({\bf q})$ is the el-ph matrix element 
defined by
\begin{align}
 M^\nu_{kk'}({\bf q}) \equiv  -\sum_{p>0} \langle \Psi(k') |
 \nabla v({\bf R}_p) | \Psi(k) \rangle
 \cdot {\bf e}^\nu_{\bf q}({\bf R}_p)
 e^{i{\bf q}\cdot {\bf R}_p}.
 \label{eq:M}
\end{align}

Putting Eq.~(\ref{eq:wf_edge}) to Eq.~(\ref{eq:M}),
we see that $M^\nu_{kk'}({\bf q})$ consists of 
the on-site 
$\langle \phi({\bf R}_i)|\nabla v({\bf R}_p)| \phi({\bf R}_i) \rangle$
and off-site
$\langle \phi({\bf R}_p)|\nabla v({\bf R}_p)| \phi({\bf R}_i) \rangle$
atomic deformation potentials (${\bf R}_p\ne {\bf R}_i$).
The off-site atomic deformation potential does not contribute to
$M^\nu_{kk'}({\bf q})$ because
$\langle \phi({\bf R}_p)|\nabla v({\bf R}_p)| \phi({\bf R}_i) \rangle$
vanishes for 
$|{\bf R}_{p\in A}-{\bf R}_i|\agt |{\bf a}_1|$.~\cite{jiang05}
We note that Eq.~(\ref{eq:M}) includes 
the effect of boundary.
To show this,
we illustrate several carbon atoms (${\bf R}_{p>0}$) 
near the zigzag edge and a fictitious atom (${\bf R}_{p<0}$) 
in Fig.~\ref{fig:me}(a).
The on-site deformation potential at ${\bf R}_1$ is given
mainly by the vibrations of carbon atoms 
at ${\bf R}_2$ and ${\bf R}_3$ as 
$-\sum_{p=2,3} \langle
\phi({\bf R}_1) | \nabla v({\bf R}_p) | \phi({\bf R}_1) \rangle \cdot
{\bf e}^\nu_{\bf q}({\bf R}_p) e^{i{\bf q}\cdot {\bf R}_p}$.
This on-site deformation potential would be canceled by 
the fictitious carbon atom at ${\bf R}_{-1}$
since ${\bf e}^\nu_{\bf q}({\bf R}_p)$ ($p=-1,2,3$)
points at the same direction.
Namely, the on-site deformation potential 
is enhanced at the edge.
This enhancement may be a reason why 
the tunnel current is unstable at the edge.~\cite{niimi05}

For a $(n,0)$ SWNT, $k$ for the edge states becomes discrete as
$k(i)=2\pi/3+2\pi i/n$ ($i=1,\ldots,n/3-1$) due to the periodic
boundary condition around the axis.  
We denote an edge state by the integer $i$ and 
write $M^\nu_{k(i)k(j)}({\bf q})$ as
$M^\nu_{ij}({\bf q})$ for simplicity.  
Putting $k(i)$ to $\xi(k)$, we obtain $\xi(k) \lesssim d_t/2$
where $d_t\equiv n |{\bf a}_1|/\pi$ is diameter
for the SWNT.

In Fig.~\ref{fig:me}(b) and (c),
we plot $|M^\nu_{14}({\bf q})|$ and $|M^\nu_{69}({\bf q})|$,
respectively, 
for the $(60,0)$ SWNT 
($d_t \approx 5$ nm).  
$|M_{14}^\nu({\bf q})|$ is chosen as an example 
that $\xi(1) \sim$ 22\AA~and $\xi(4) \sim$ 4.4\AA~are much longer
than the carbon-carbon bond length $a_{\rm cc}\sim$ 1.4\AA, 
while $|M_{69}^\nu({\bf q})|$ is chosen as another
example that $\xi(6) \sim 2.4$\AA~and $\xi(9) \sim 0.9$\AA~are
comparable to $a_{\rm cc}$.
As for acoustic modes, the LA mode 
couples strongly to the edge state.  
The oTA mode contributes to $|M^\nu_{14}({\bf q})|$, 
whereas the iTA mode is negligible.  
Since the LA and oTA modes change the
area of a hexagonal lattice, they contribute to
on-site deformation potential.
For optical modes, 
the iTO and LO modes are important.
$|M_{14}^{\rm iTO}({\bf q})|$ decreases with increasing $q_t$,
while $|M_{14}^{\rm LO}({\bf q})|$ increases with increasing $q_t$.
As shown in Fig.~\ref{fig:me}(c), 
the deformation potential is stronger for the smaller
localization length.  
The behavior of the iTO and LO modes
is due to the boundary deformation potential.
To prove this, 
we show in the inset,
the matrix element without the boundary,
which is defined by Eq.~(\ref{eq:M}) 
including (fictitious) carbon atoms at $p<0$ in Fig.~\ref{fig:me}(a). 
The boundary deformation potential
depends on the direction of ${\bf e}_{\bf q}^{\nu}$
and the magnitude 
is maximum when ${\bf e}_{\bf q}^{\nu}$ is parallel to ${\bf T}$.
${\bf e}_{\bf q}^{\rm iTO}$ (${\bf e}_{\bf q}^{\rm LO}$)
has a large element parallel to ${\bf T}$
when $q_t < \sqrt{3}q$ 
($q_t > \sqrt{3}q$)
as shown as a vertical line in Fig.~\ref{fig:me}(b) and (c).
On the other hand, 
${\bf e}_{\bf q}^{\rm oTO}$ is perpendicular to the SWNT axis 
(or parallel to ${\bf a}_1 \times {\bf T}$)
and the boundary effect of the oTO mode does not appear
in $|M_{ij}^{\rm oTO}({\bf q})|$.

Now we apply $|\alpha_{kk'}^\nu({\bf q})|$ 
to the Eliashberg equation.
The Eliashberg equation
includes the effects of phonon retardation and electron self-energy,
which are not taken into account in the BCS
theory.~\cite{eliashberg60,nambu60} 
The phonon retardation is included 
by the Matsubara frequency:
$\omega_n= k_{\rm B} T (2n+1)\pi$ where 
$n$ is integer and $|\omega_n| \le \omega_{\rm D}$ where
$\omega_{\rm D} = 0.2$ eV is the Debye energy.~\cite{saito98book}
Since the gap function,
$\Delta(k,i\omega_n)$, 
vanishes at $T_{\rm c}$,
the Eliashberg equation can be linearized 
at $T_{\rm c}$ to get the gap equation:
\begin{align}
 \Delta(k,i\omega_n) 
 &= \frac{2k_{\rm B} T_{\rm c}}{N_{\rm u}} 
 \sum_{k',m,q_t,\nu}
 \frac{|\alpha^\nu_{kk'}({\bf q})|^2 \omega_\nu({\bf q})}{(\omega_n -
 \omega_m)^2 + 
 \omega_\nu^2({\bf q})} \nn \\
 &\times
 |G(k',i\omega_m)|^2 \Delta(k',i\omega_m),
 \label{eq:gap_eq}
\end{align}
where $G(k,i\omega_n)$ is a thermal Green function of electron,
$G(k,i\omega_n)=
(i\omega_n-(E(k)-E_{\rm F})-\Sigma(k,i\omega_n))^{-1}$.
Here, $\Sigma(k,i\omega_n)$ is the self-energy, which 
is determined self-consistently by
\begin{align}
 \Sigma(k,i\omega_n)
 = \frac{2 k_{\rm B} T_{\rm c}}{N_{\rm u}} 
 \sum_{k',m,q_t,\nu}
 \frac{|\alpha^\nu_{kk'}({\bf q})|^2 \omega_\nu({\bf
 q})}{(\omega_n - \omega_m)^2 + \omega_\nu^2({\bf q})}G(k',i\omega_m).
 \label{eq:self-ene}
\end{align}
After calculating $\Sigma(k,i\omega_n)$ in Eq.~(\ref{eq:self-ene}),
we solve Eq.~(\ref{eq:gap_eq}).

\begin{figure}[htbp]
 \begin{center}
  \includegraphics[scale=0.45]{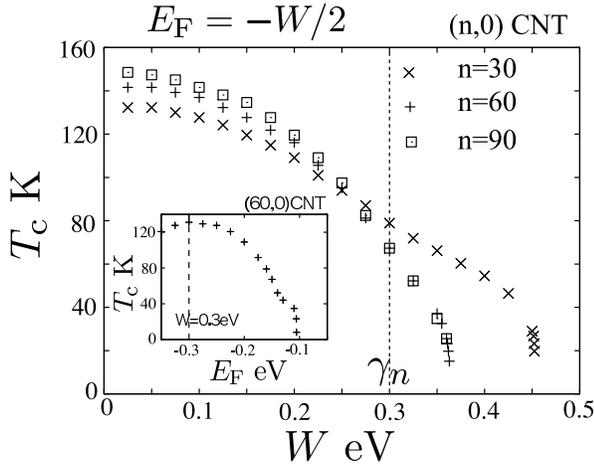}
 \end{center}
 \caption{The dependence of $T_{\rm c}$ on $W$ is plotted
 for $(30,0)$, $(60,0)$ and $(90,0)$ zigzag SWNTs.
 The Fermi energy is assumed to be located at the center of 
 the band ($E_{\rm F} = -W/2$).
 The nnn hopping gives $W= 0.3$ eV.
 (inset)
 The dependence of $T_{\rm c}$ on $E_{\rm F}$
 for the $(60,0)$ SWNT with $W=\gamma_n$.
 } 
 \label{fig:tc}
\end{figure}

In Fig.~\ref{fig:tc}, we show $T_{\rm c}$ as a function of $W$ 
for $(n,0)$ SWNTs with $n=30,60,$ and 90,
where we assume $E_{\rm F}=-W/2$.
$T_{\rm c}$ decreases with increasing $W$ 
and $T_{\rm c}$ vanishes at critical values, $W_{\rm c}$.
The increase of $W$ corresponds to 
the decrease of the LDOS around the Fermi energy.
The values of $W_{\rm c}$ are 0.46 eV and 0.37 eV, respectively for 
$n=30$ and $n \ge 60$. 
Those values of $W_{\rm c}$ are close to $\gamma_n$
(the dashed line in Fig.~\ref{fig:tc}).
When $n$ ($d_t$) is relatively small,
all edge states couple strongly to
the boundary deformation potential
since ${\rm max}(\xi) \lesssim d_t/2$.
The strong el-ph coupling for $n=30$
makes $W_{\rm c}$ larger than that for $n \ge 60$.
It is also noted that
excluding optical modes
makes $T_{\rm c}$ and $W_{\rm c}$ both smaller.
In this case, we obtain $T_{\rm c} \sim$ 70 K at $W=0$ eV
and $W_{\rm c} \sim $ 0.21 eV for $n=60$.
Although the values of $A^\nu({\bf q})$ for optical modes are smaller
than those of acoustic modes,  
the iTO and LO modes contribute to 
Eqs.~(\ref{eq:gap_eq}) and (\ref{eq:self-ene}) because of the
large values of $|M_{ij}^\nu({\bf q})|$ 
due to the boundary deformation potential. 
A large value of $n$ corresponds to the zigzag edge of graphene.
Remarkably, $T_{\rm c}$ for $n=120$ 
has a curve quite similar to $T_{\rm c}$ for $n=90$.
This suggests that $T_{\rm c}$ converges and 
$n = 90$ is large enough to represent a graphene.

It is important to note that 
the calculated $T_{\rm c}$ is sensitive 
to the energy position of $E_{\rm F}$. 
We plot $T_{\rm c}$ as a function of $E_{\rm F}$
for a $(60,0)$ SWNT with $W=\gamma_n$
in the inset of Fig.~\ref{fig:tc}.
When $E_{\rm F}$ exists at the top of the
energy band of the edge states, $T_{\rm c}$ becomes less than 1 K.
When $E_{\rm F} \sim -0.1$ eV, $T_{\rm c}$ decreases rapidly
since the inelastic scattering process 
is suppressed by the absence of the scattered state.
This is a reason why
$T_{\rm c}$ is sensitive to the position of $E_{\rm F}$.
We also calculated $T_{\rm c}$ for extended states 
around the Fermi energy of 
the $(60,0)$ SWNT using the Eliashberg equation.
The calculated $T_{\rm c}$ is less than 0.1 K. 
Thus the extended states do not contribute to $T_{\rm c}$.

The observed $T_{\rm c}$ should be smaller than our estimation.  
In fact, a lattice defect along the edge decreases LDOS 
and reduces $T_{\rm c}$. 
The Coulomb repulsive interaction might decrease $T_{\rm c}$, too.
Fujita {\it et al}. showed that 
the edge states develop a local ferro-magnetism 
in the presence of a large Hubbard $U$ 
comparable to $W$.~\cite{fujita96} 
Since the edge states are localized at the edge,
they might have a quantum fluctuation intrinsic to 
1D system.
The Tomonaga-Luttinger liquid theory 
may be suitable to calculate the 
correlation function.

In summary, using the Eliashberg equation,
we clarify that $W$ and $E_{\rm F}$ position 
is sensitive to $T_{\rm c}$ of the edge states in SWNTs and graphene. 
The rather high value of $T_{\rm c}$ obtained is a result of
LDOS enhancement by the edge states, and
the on-site and boundary deformation potentials of
the el-ph interaction for the edge states.
If nanotube superconductivity is given by el-ph interaction,
the edge-state superconductivity is a unique candidate
since $T_{\rm c}$ of the bulk is negligible.
Edge (surface) state superconductivity is potentially 
a key concept for designing superconductors on the nanometer scale.

R. S. acknowledges a Grant-in-Aid (No. 16076201) from MEXT.


\end{document}